\documentclass[12pt]{article}
\usepackage{a4wide}
\usepackage{amssymb}
\usepackage{amsmath}
\usepackage{amsthm}
\usepackage{graphicx}
\usepackage{mathdots}
\usepackage{slashed}
\usepackage{verbatim}
\usepackage{xcolor}
\usepackage{dsfont}

\usepackage{tikz}

\usepackage{hyperref}

\usepackage{IEEEtrantools}

\usepackage{ifpdf}

\ifpdf
  \usepackage{pdflscape}
\else
  \usepackage{lscape}
\fi

\DeclareMathOperator{\U}{U}
\DeclareMathOperator{\SL}{SL}

\DeclareMathOperator{\osp}{osp}

\DeclareMathOperator{\vol}{vol}
\DeclareMathOperator{\AdS}{AdS}

\newcommand{\bC}{\mathds{C}}
\newcommand{\bR}{\mathds{R}}
\newcommand{\bH}{\mathds{H}}

\newcommand{\bI}{\mathds{1}}

\newcommand{\tCR}{\partial_{b}}
\newcommand{\tCRc}{\bar{\partial}_{b}}

\newcommand{\ket}[1]{\left\vert #1 \right\rangle}

\begin{document}

\begin{titlepage}

\vskip 2cm

\begin{center}
{\Large \bfseries
Superconformal Symmetry in the Kaluza-Klein Spectrum of Warped AdS(3)
}

\vskip 1.2cm

Johannes Schmude and Orestis Vasilakis\textsuperscript{1}

\bigskip
\bigskip

\begin{tabular}{c}
Department of Physics, Universidad de Oviedo, \\
Avda.~Calvo Sotelo 18, 33007, Oviedo, Spain\\
\end{tabular}

\vskip 1.5cm

\textbf{Abstract}
\end{center}

\medskip
\noindent
We study the Kaluza-Klein spectrum of warped $\AdS_3$ compactifications of type IIB with five-form flux which are dual to $\mathcal{N}=(0,2)$ SCFTs in two dimensions. We prove that the spectra of fluctuations of both the spin 2 sector of the graviton and the axio-dilaton are bounded. At the bound the modes have the correct quantum numbers to be chiral primaries and descendants thereof respectively. Moreover, we prove that the same modes give rise to superpartners in the dilatino spectrum. 
Our results show that a subset of the mesonic chiral ring of the dual SCFT is isomorphic to the first Kohn-Rossi cohomology groups. As an example, we consider the compactification of four-dimensional $Y^{p,q}$ theories on Riemann surfaces for the case of the universal twist. We conclude by studying fluctuations of the three-form, where we are able to identify Betti multiplets after imposing some mild assumptions.

\bigskip
\vfill

\footnotetext[1]{schmudejohannes@uniovi.es,vasilakisorestis@uniovi.es}

\end{titlepage}
\tableofcontents

\section{Introduction \& Summary}

In the supergravity limit, the AdS/CFT dictionary maps operators of the boundary CFT to supergravity fields in the bulk. It follows that one can infer the operator content of the CFT from the Kaluza-Klein spectrum of the dual geometry. For a large class of backgrounds this has been fully solved. These backgrounds are usually of Freund-Rubin type and the internal manifold has a coset structure. While the calculations are often quite involved, one can essentially use group theoretic methods to calculate the full spectrum \cite{Kim:1985ez,Ceresole:1999ht,Ceresole:1999zs,Deger:1998nm,deBoer:1998kjm,Merlatti:2000ed,Fabbri:1999mk}.

When either of these simplifications ceases to apply, the situation becomes more involved. The well-known $\AdS_5$ and $\AdS_4$ compactifications on Sasaki-Einstein five and seven manifolds respectively are still of Freund-Rubin type, yet in the generic case the internal manifold no longer carries a coset structure. However, since the dual field theory is still a superconformal field theory, it is clear that the KK spectrum has to respect the constraints imposed by superconformal symmetry. That is, the spectrum has to fit into superconformal multiplets with the individual modes satisfying the bounds imposed by unitarity. In \cite{Eager:2012hx,Eager:2013mua,Schmude:2013dua} it was shown that the Sasaki-Einstein structure of the background is sufficient to arrange fluctuations around a generic solution of this type into multiplets and identify short multiplets with certain cohomology groups, thus allowing the calculation of the superconformal index.

The situation gets more intricate when the geometry is not of Freund-Rubin type. Here, a number of individual results are known such as \cite{Klebanov:2009kp,Ahn:2009et,Bachas:2011xa,Passias:2016fkm,Pang:2015rwd}. To understand the various complications, we should briefly sketch the general procedure. In order to obtain the Kaluza-Klein spectrum, one has to first calculate the \emph{mass operators}. That is if we consider a fluctuation of some field $B$,
\begin{equation}
  B (\AdS_{d+1}, M) = B(\AdS_{d+1}) \otimes Y(M),
\end{equation}
the linearised equations of motion for the \emph{fluctuation} $B(\AdS_{d+1}, M)$ will yield some differential equation for $B(\AdS_{d+1})$ and $Y(M)$ from which the mass of the \emph{mode} $B(\AdS_{d+1})$ follows in terms of the \emph{wave function} $Y(M)$. While the mass operators are fairly simple to derive in the Freund-Rubin case,\footnote{
  A detailed derivation of the mass operators of Freund-Rubin compactifications is given in \cite{Larsson2004}. The essentials of harmonic decomposition are nicely summarized in \cite{Bailin:1987jd}. Another very readable review of Kaluza-Klein theory is \cite{Duff:1986hr}. 
}
the calculations can get involved for arbitrary backgrounds. The usual complications are the presence of an additional warp factor as well as flux that is not just proportional to a volume form.

Nevertheless, the solution of the KK problem for more general backgrounds is of obvious interest. In this paper we will make some progress into this direction by studying a specific class of 1/8 BPS $\AdS_3$ compactifications of type IIB that are dual to two-dimensional $\mathcal{N}=(0,2)$ SCFTs with $\U(1)$ $R$-symmetry. These are generally warped and carry five-form flux, yet the remaining fields are trivial. A first classification was made in \cite{Kim:2005ez}. For subsequent work refining and extending these results see \cite{Gauntlett:2006af,Gauntlett:2006ns,Gauntlett:2006qw,Gauntlett:2007ph,Gauntlett:2007ts}.

The backgrounds are of the form $\AdS_3 \times_A M_7$ with $M_7$ a one-dimensional foliation over a six manifold that is conformally K\"ahler. In other words, the general form of the metric is
\begin{IEEEeqnarray}{rCl}\label{eq:IIB_metric}
  ds_{10A}^2 &=& L^2 e^{2A} \left[ ds^2(\AdS_3) + ds_7^2 \right], \nonumber\\
  ds_7^2 &=& e^{-4A} ds_6^2 + \frac{1}{4} (dz+P)^2,
\end{IEEEeqnarray}
and $ds_6^2$ is K\"ahler. The curvature of the six manifold satisfies the non-trivial constraint
\begin{equation}\label{eq:M6_curvature_constraint}
  \Delta_6 R = \frac{1}{2} R^2 - R^{mn} R_{mn},
\end{equation}
where $\Delta_6$ is the Laplacian on $M_6$. The warp factor is fully determined in terms of the Ricci scalar on $M_6$, $e^{-4A} = \frac{1}{8} R$. The only non-trivial field is the five-form flux:
\begin{IEEEeqnarray}{rCl}\label{eq:F5}
  F_5 &=& L^4 (1 + \star_{10A}) \vol(\AdS_3) \wedge F, \nonumber\\
  F &=& \frac{n}{2} J - \frac{1}{8} d\left[ e^{4A} (dz + P) \right],
\end{IEEEeqnarray}
where $J$ is the K\"ahler form on $M_6$.
We introduce the one-form $\eta = \frac{dz+P}{2}$ and note that $d\eta = \frac{n}{2} \rho$, the Ricci-form of $M_6$. Similarly to the notation that is familiar from the Sasaki-Einstein case \cite{Martelli:2006yb,Sparks:2010sn}, we denote the dual vector as $\xi = 2 \frac{\partial}{\partial z}$. The constant $n$ satisfies $n = \pm 1$. In the remainder of the paper we will set the $\AdS_3$ curvature radius $L$ to one.

There is an interesting complication regarding the realization of the $\U(1)$ $R$-symmetry on the string theory side of the duality that has been discovered in \cite{Benini:2015bwz}. There, the authors studied flows from four-dimensional $\mathcal{N}=1$ theories to two-dimensional ones via twisted compactification on Riemann surfaces. Using $c$-extremization \cite{Benini:2013cda}, the IR $R$-symmetries of the two-dimensional theories were found to not only receive contributions from the $R$- and flavor-symmetries of the four-dimensional theories but also from baryonic symmetries. It follows that the $R$-symmetry of the two-dimensional theory cannot in general be realized as an isometry of the background; instead it consists of an isometry and a ``baryonic contribution'' to the $R$-symmetry. To understand the consequences for the study of Kaluza-Klein spectra, let us consider fluctuations around $\AdS_5$ geometries in type IIB. These are dual to mesonic operators and thus not charged under baryonic symmetries. Returning to the question of Kaluza-Klein fluctuations around \eqref{eq:IIB_metric}, analogy suggests that supergravity fluctuations should be insensitive to the ``baryonic contribution'' to the $R$-symmetry. In other words, for the purpose of Kaluza-Klein analysis it is sufficient to consider the part of the $R$-symmetry that is generated by an isometry, which in the case at hand is known to be the Killing vector $\xi$ \cite{Kim:2005ez,Donos:2008hd}. Our results will justify this treatment.

The comparison between \eqref{eq:IIB_metric} and the Sasaki-Einstein case is quite fitting. The seven manifold is actually Cauchy-Riemann (CR). That is, the tangent bundle splits as
\begin{equation}\label{eq:M7_Tangent_bundle_decomposition}
  T_{\bC}M_7 = T^{1,0} \oplus T^{0,1} \oplus \bC \xi
\end{equation}
and the distribution $T^{1,0}$ is integrable, $[T^{1,0}, T^{1,0}] \subseteq T^{1,0}$. Note that all Sasakian manifolds are CR. Just as Sasaki-Einstein manifolds can be defined as the base of a Calabi-Yau cone, CR-manifolds can be considered as the boundary of some (possibly singular) complex variety \cite{harvey1974boundaries,harvey1975boundaries,yau1981kohn}. This was also shown in \cite{Gauntlett:2007ts}, where eight-dimensional conical, complex geometries that reduce to the above system \eqref{eq:IIB_metric} were explicitly constructed. As an illustration, we note that the integrability condition satisfied by the tangent bundle of the CR manifold follows directly from the familiar integrability condition satisfied by $T_{\bC} M_8 = T^{1,0}_8 \oplus T^{0,1}_8$.

Our strategy is then to generalize the methodology of the Sasaki-Einstein case \cite{Eager:2012hx,Eager:2013mua,Schmude:2013dua} to the case at hand. That is, there has to be a geometric equivalent of the unitarity bounds satisfied by the two-dimensional $\mathcal{N}=2$ algebra \cite{Bars:1982ep,Gunaydin:1986fe,Lerche:1989uy,Adams:2005tc}. Moreover, the bound should be intimately related to the CR structure of the seven manifold and modes at the bound should define some cohomology group reflecting the structure of the chiral ring. Finally, the supergravity spectrum has to respect superconformal symmetry from which it follows that the spectra of the various mass operators should be related. For Freund-Rubin compactifications over coset spaces this has been discussed in \cite{DAuria:1984vy}.\footnote{See also \cite{Pope:1982ad} for a discussion of mappings between the spectra of various differential operators on K\"ahler manifolds.
}
Turning to more explicit calculations, it is known that the differential equations governing the wave functions often simplify considerably at the unitarity bound. This was used in \cite{Ardehali:2013xla} to calculate the scalar spectrum of $Y^{p,q}$ manifolds \cite{Gauntlett:2004yd} explicitly following earlier work \cite{Berenstein:2005xa,Kihara:2005nt,Oota:2005mr}. Finally, the spectrum of the $\AdS_3 \times S^3 \times T^4$ and $\AdS_3 \times S^3 \times K3$ geometries dual to the near horizon region of a D1-D5 intersection has been studied extensively in the early days of the AdS/CFT correspondence \cite{Deger:1998nm,deBoer:1998kjm,Maldacena:1998bw,deBoer:1998us,Aharony:1999ti}. While these carry three-form instead of five-form flux and thus do not fall into the class of \cite{Kim:2005ez} considered here, the T-dual D3-D3 intersection with near horizon geometry $\AdS_3 \times S^3 \times T^4$ and five-form flux does \cite{Behrndt:1996pm,Gauntlett:1996pb}.

Returning to the geometries at hand and the CR structure on the internal seven manifold, we note that the decomposition \eqref{eq:M7_Tangent_bundle_decomposition} also extends to the cotangent bundle,
\begin{equation}
  T^*_{\bC}M_7 = \Omega^{1,0} \oplus \Omega^{0,1} \oplus \bC \eta.
\end{equation}
Since the distribution is integrable, it follows that the exterior differential can be decomposed in the same way,
\begin{equation}\label{eq:decomposition_exterior_differential}
  d = \tCR + \tCRc + \eta \wedge \pounds_\xi
\end{equation}
and that $\tCRc^2 = \tCR^2 = 0$. That is, $\tCRc$ is the $(0,1)$ component of $d$. It is referred to as the \emph{tangential Cauchy-Riemann operator}. The situation proceeds in analogy to the case of complex geometry. The sequence
\begin{equation}
  \hdots \xrightarrow{\tCRc} \Omega^{p,q-1} \xrightarrow{\tCRc} \Omega^{p,q}, \xrightarrow{\tCRc} \Omega^{p,q+1} \xrightarrow{\tCRc} \hdots
\end{equation}
is exact and allows us to define the \emph{Kohn-Rossi cohomology groups} $H_{\tCRc}^{p,q}(M_7)$. In the Sasaki-Einstein case, the wave functions saturating the unitarity bound are holomorphic in the sense of the tangential Cauchy-Riemann operator\footnote{In this paper we will usually use the term holomorphic in this sense -- i.e.~to denote $\tCRc Y = 0$.
}
and short multiplets correspond thus to equivalence classes in Kohn-Rossi cohomology. There are equivalent statements on the Calabi-Yau cone \cite{Eager:2012hx,Eager:2013mua,Eager:2015hwa}.

\begin{figure}
\begin{center}
\begin{tikzpicture}
\draw[<->] (6,0) node[below]{$R$} -- (0,0) --
    (0,6) node[left]{$\bar{h}$};
\draw [dashed] (4,1) -- (6,1);
\draw [red,thick] (4,3) -- (6,3);
\draw (4,5) -- (6,5);
\draw [blue,thick] (1,2) -- (3,2);
\draw [blue,thick] (1,4) -- (3,4);
\draw (1,6) -- (3,6);
\draw [dotted,<-] (3.25,2.25) -- (3.75,2.75);
\draw [dotted,<-] (3.25,3.75) -- (3.75,3.25);
\draw [dotted,->] (5,2) -- (5,2.5);
\draw [dotted,->] (5,4) -- (5,4.5);
\draw [dotted,->] (5,5.5) -- (5,6);
\draw [dotted,->] (2,3) -- (2,3.5);
\draw [dotted,->] (2,5) -- (2,5.5);
\node [above] at (5,3) {$\ket{\frac{E_0+2}{2}, q}_0$};
\node [above] at (2,2) {$\ket{\frac{E_0+1}{2}, q-1}_{\frac{1}{2}}$};
\node [above] at (2,4) {$\ket{\frac{E_0+3}{2}, q-1}_{-\frac{1}{2}}$};
\node [above] at (5,1) {$\ket{\frac{E_0}{2}, q}$};
\end{tikzpicture}
\end{center}
\caption{A summary of our results for the axio-dilaton as well as the dilatino. The figure only includes the right handed quantum numbers $\ket{\bar{h}, q}_s$ since $h$ follows directly from the helicity via $h - \bar{h} = s$. The action of $\bar{L}_{-1}$ and $G^-_{\pm 1/2}$ is indicated with dotted arrows. The red mode comes from the axio-dilaton and is discussed in section \ref{sec:minimally_coupled_scalar}, the blue modes are the dilatino modes discussed in section \ref{sec:dilatino_analysis}. At the unitarity bound we have $E_0 = q$. The dashed mode is the chiral primary.}
\label{fig:axio-dilaton_dilatino_summary}
\end{figure}
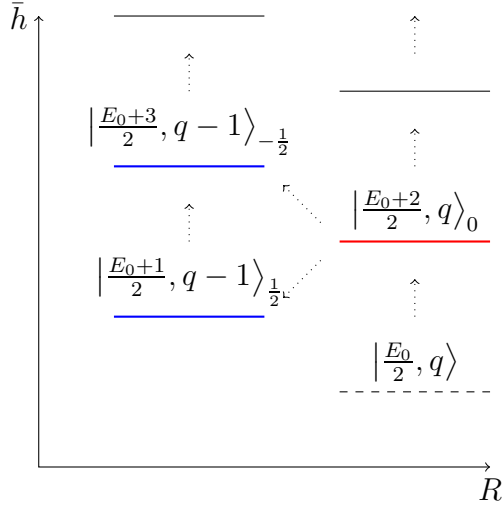

We will obtain similar results for spin 2 modes in the graviton spectrum and general fluctuations of the axio-dilaton. In section \ref{sec:minimally_coupled_scalar} we prove that the mass operator governing these is bounded and that the bound is saturated if the wave function is holomorphic. Moreover, the states have the correct quantum numbers to be chiral primaries or descendants thereof respectively, showing that a subset of the chiral ring is isomorphic to $H^{0,0}_{\tCRc}(M_7)$. In section \ref{sec:dilatino_analysis} we find that any wave function appearing in the spectrum of the axio-dilaton -- holomorphic or not -- gives rise to two fluctuations of the dilatino which have the correct quantum numbers to be superpartners of the axio-dilaton mode. These results are summarized in figure \ref{fig:axio-dilaton_dilatino_summary}.

An obvious question is whether we can generalize these successes away from the comparatively simple spin 2, axio-dilaton and dilatino fluctuations to the remainder of the spectrum. With this in mind we take a first look at the three-form equation in section \ref{sec:three-form}. The situation is involved, yet under some mild assumptions we are able to identify Betti multiplets; that is, multiplets arising from non-trivial de Rham cohomology groups $H^1(M_7)$ and $H^2(M_7)$. Our analysis shows that the deformation of the Laplace operator seen in the case of the axio-dilaton can be generalized suitably to $p$-forms.

As we mentioned previously, a particularly interesting class of the geometries in question has recently been constructed in \cite{Benini:2015bwz} via twisted compactification of four-dimensional $\mathcal{N}=1$ quiver gauge theories dual to type IIB on $\AdS_5 \times Y^{p,q}$. Our results on the spectrum together with those of \cite{Eager:2012hx,Eager:2013mua,Ardehali:2013xla} lead to an interesting question: If the short multiplets of the four-dimensional theory correspond to certain holomorphic sections on the Calabi-Yau cone over $Y^{p,q}$, and if at least some of the short multiplets of the two-dimensional theory correspond to holomorphic sections of the cone over $M_7$, how are the two spectra related? Our methods are suitable to address this question and we will give a first glimpse of this in section \ref{Sec:ExampleUnivFlow}.

\section{The Minimally Coupled Scalar and the Unitarity Bound}
\label{sec:minimally_coupled_scalar}

Consider fluctuations of the axio-dilaton. These yield a complex, harmonic scalar in the ten-dimensional warped geometry. Using the explicit form of the background \eqref{eq:IIB_metric}, the linearised equation of motion can be rewritten as
\begin{equation}
  0 = \Delta_{10A} B = e^{-2A} \left( \Delta_3 + \Delta_7 - 8 g_7^{\kappa\lambda} \partial_\kappa A \partial_\lambda \right) B.
\end{equation}
Here, $\Delta_3$ and $\Delta_7$ are the Laplace operators on $\AdS_3$ and $M_7$ respectively. Thus, we need to study the spectrum of the operator
\begin{equation}\label{eq:minimally_coupled_scalar}
  \mathcal{L}_0 = \Delta_7 - 8 g_7^{\kappa\lambda} \partial_\kappa A \partial_\lambda = - e^{-8A} \nabla_\kappa ( e^{8A} g_7^{\kappa\lambda} \partial_\lambda).
\end{equation}
As a matter of fact, $\mathcal{L}_0$ doubles also as the mass operator of spin 2 modes in the graviton spectrum. This has been previously discussed in \cite{Klebanov:2009kp,Bachas:2011xa} and we give a brief summary of the derivation in appendix \ref{sec:appendix_spin_2}. In an abuse of notation we will often use the same symbol for a differential operator and its eigenvalues.

To begin, note that the spectrum of $\mathcal{L}_0$ is positive definite. Consider a generic eigenfunction $Y_0$ with $\mathcal{L}_0 Y_0 = \delta_0 Y_0$. Then, $\delta_0 \geq 0$ follows from integration by parts
\begin{equation}
  \delta_0 \int e^{8A} \sqrt{g_7} \vert Y_0 \vert^2 = \int e^{8A} \sqrt{g_7} \bar{Y}_0 \mathcal{L}_0 Y_0 
  = \int e^{8A} \sqrt{g_7} \vert \partial Y_0 \vert^2 \geq 0.
\end{equation}
The main ``trick'' we have used here is to sneak a factor of $e^{8A}$ into the integral. Similarly, the functions $Y_0$ appear as an orthonormal system with respect to the inner product defined by $(X_0, Y_0)_8 = \int_{M_7} e^{8A} \sqrt{g}_7 \bar{X}_0 Y_0$.\footnote{We would like to thank Diego Rodriguez-Gomez for discussions that led to this observation.}
This modified inner product will be very useful when considering more complicated cases of higher rank forms in section \ref{sec:three-form}.

To proceed, we introduce local, complex coordinates $w^i$ on $M_6$ and define the operators
\begin{equation}\label{eq:tCR_on_scalars}
  \tCR = dw^i \left( \frac{\partial}{\partial w^i} - P_i \frac{\partial}{\partial z} \right), \qquad
  \tCRc = d\bar{w}^{\bar{j}} \left( \frac{\partial}{\partial \bar{w}^{\bar{j}}} - P_{\bar{j}} \frac{\partial}{\partial z} \right).
\end{equation}
This definition is equivalent to that given in the introduction. Integrating by parts one finds that
\begin{IEEEeqnarray}{rCl}
  \int e^{8A} \sqrt{g_7} \bar{Y}_0 \mathcal{L}_0 Y_0 &=& \int e^{8A} \sqrt{g_7} \left[ e^{4A} g_6^{i\bar{j}} \left( 2 \partial_{bi} \bar{Y}_0 \bar{\partial}_{b\bar{j}} Y_0 + \bar{Y}_0 \left\lbrack \nabla_{bi}, \bar{\nabla}_{b\bar{j}} \right\rbrack Y_0 \right) - 4 \bar{Y}_0 \partial_z^2 Y_0 \right] \nonumber\\
  &=& \int e^{8A} \sqrt{g_7} \left[ 2 e^{4A} \vert \tCRc Y_0 \vert^2 - 4 \bar{Y}_0 (\partial_z^2 Y_0 + \imath n \partial_z Y_0 ) \right].
\end{IEEEeqnarray}
Here, $\nabla_{bi}$ denotes the Levi-Civita connection on $M_6$ twisted as in \eqref{eq:tCR_on_scalars} and at the last step made use of the equation for the warp factor:
\begin{equation}
  g_6^{i\bar{j}} \left\lbrack \nabla_{bi}, \bar{\nabla}_{b\bar{j}} \right\rbrack Y_0 = - n g_6^{i\bar{j}} \rho_{i\bar{j}} \partial_z Y_0 = -4 \imath n e^{-4A} \partial_z Y_0.
\end{equation}

We restrict to $n = +1$. The discussion for $n = -1$ is analogous. Since $\xi$ is Killing, we can diagonalize $\mathcal{L}_0$ and $\pounds_\xi$ simultaneously. If $Y_0$ satisfies $\pounds_\xi Y_0 = \imath q Y_0$, it follows that
\begin{equation}\label{eq:unitarity-bound_axio-dilaton}
  \mathcal{L}_0 \equiv E_0^2 + 2 E_0 \geq q^2 + 2q \, ,
\end{equation}
where as it is customary we introduced the notation $\mathcal{L}_0 \equiv E_0^2 + 2 E_0$ for the $\mathcal{L}_0$ eigenvalue. Thus the bound (\ref{eq:unitarity-bound_axio-dilaton}) is $E_0 \geq q$. It is saturated if and only if $Y_0$ is holomorphic in the sense of $\tCRc Y_0 = 0$.

Let us map this to a dual operator. In general, an operator in the spectrum of $(0, 2)$ theories is of the form
\begin{equation}
  \vert h \rangle \otimes \vert \bar{h}, q \rangle.
\end{equation}
Multiplets are obtained by repeated application of $L_0$, $L_{\pm 1}$ in the left handed sector and $\bar{L}_0$, $\bar{L}_{\pm 1}$ and $G^\pm_{\pm 1/2}$ in the right handed sector. The $R$-symmetry current is $J_0$. The states are labelled according to the eigenvalues of the operators $L_0$, $\bar{L}_0$ and $J_0$. That is, $L_0 \ket{h} = h \ket{h}$, $\bar{L}_0 \ket{\bar{h}, q} = \bar{h} \ket{\bar{h}, q}$ and $J_0 \ket{\bar{h}, q} = q \ket{\bar{h}, q}$. Starting with fluctuations of the graviton we have $h + \bar{h} = \Delta_\pm$ and $h - \bar{h} = \pm 2$ while the relevant mass-dimension formula is $\Delta_\pm = 1 \pm \sqrt{1 + m^2}$. For the helicity $h - \bar{h} = +2$ we obtain the operator
\begin{equation}\label{eq:spin_2_fluctuation_operator}
  \ket{\frac{E_0 + 4}{2}} \otimes \ket{\frac{E_0}{2}, q}.
\end{equation}
At the bound, $2 \bar{h} = E_0 = q$ and it is clear that we have found a chiral primary. We will thus refer to \eqref{eq:unitarity-bound_axio-dilaton} as the unitarity bound. For the special case of a constant wave function, the dual operator in the graviton spectrum is the energy momentum tensor. In the above we have tacitly assumed that $q \geq 0$. For $q \leq 0$ one would consider anti-holomorphic wave functions.

Next we turn to the axio-dilaton. Since $B$ is a scalar fluctuation, we have $h = \bar{h}$ and thus $2 \bar{h} = \Delta_+ = E_0 + 2$ where we used the mass-dimension formula for scalars in two dimensions, $\Delta_\pm = 1 \pm \sqrt{1 + m^2}$. In other words, the operator in question is
\begin{equation}\label{eq:axio-dilaton_fluctuation_operator}
  \ket{\frac{E_0+2}{2}} \otimes \ket{\frac{E_0+2}{2},q}.
\end{equation}
At the bound it takes the form
\begin{equation}\label{eq:scalar_Delta_at_bound}
  \ket{\frac{q+2}{2}} \otimes \ket{\frac{q+2}{2},q}.
\end{equation}
This mode is not a chiral primary for which $\bar{h} = \frac{\vert q\vert}{2}$, yet since $Y_0$ satisfies a differential condition and saturates a bound one should expect it to be a descendant of one. We will return to this point in section \ref{sec:superconformal_algebra}.

Equation \eqref{eq:scalar_Delta_at_bound} is not the only solution to the mass-dimension formula. Indeed, it might appear that $2\bar{h} = \Delta_- = - q$ saturates the unitarity bound if the $R$-charge is negative.\footnote{Unitarity of the $\mathcal{N}=1$ algebra imposes $\bar{h} \ge 0$ in the NS sector.
}
However, this possibility does not agree with our findings in section \ref{sec:dilatino_analysis} or with the structure of short representations of the $\mathcal{N}=2$ algebra. We will give several reasons for this: In section \ref{sec:dilatino_analysis} we will show that any eigenmode of the operator $\mathcal{L}_0$ gives rise to two eigenmodes in the spectrum of the dilatino. The situation is summarized in figure \ref{fig:axio-dilaton_dilatino_summary}. Here, $\ket{\frac{E_0+2}{2},q}_0$ is the mode discussed in this section, yet clearly it is not of lowest weight when compared with the dilatino modes. Furthermore, it follows from the $\mathcal{N}=2$ algebra that a chiral primary operator satisfying $\bar{h} = - \frac{q}{2}$ is annihilated by the supercharge $G^-_{-1/2}$. Yet $G^-_{-1/2} \ket{\frac{E_0+2}{2},q}_0 = \ket{\frac{E_0+3}{2},q-1}_{-\frac{1}{2}}$ which is generally not zero, resulting in another contradiction. Independently of this argument one would expect that all holomorphic modes correspond to modes with $R$-charges of the same sign. Finally, solutions with negative R-charge would correspond to meromorphic rather than holomorphic functions which would lead to non-normalizable modes. In light of all the above we reject the solution $\Delta_-$.

Summarizing, we found that for every element $Y_0$ of the Kohn-Rossi cohomology group $H_{\tCRc}^{0,0}$ there is a short superconformal multiplet including the mode \eqref{eq:spin_2_fluctuation_operator} with $E_0 = q$. The axio-dilaton fluctuations \eqref{eq:scalar_Delta_at_bound} might lie in the same multiplet. If they do not, every element of $H_{\tCRc}^{0,0}$ will give rise to a second superconformal multiplet. Since the bound \eqref{eq:unitarity-bound_axio-dilaton} is saturated, one would expect these to be short as well. As remarked earlier, the group $H_{\tCRc}^{0,0}(M_7)$ can be lifted to $H_{\bar{\partial}}^{0,0}(C(M_7))$. Thus one can count these short multiplets by counting holomorphic functions on the variety $C(M_7)$ \cite{yau1981kohn,Gauntlett:2007ts,Eager:2012hx}.

\section{Examples}
\label{sec:Examples}

\subsection{The D3-D3 Intersection}
\label{sec:Example_D3D3}

The probably simplest solution in the class \eqref{eq:IIB_metric} is the $\AdS_3 \times S^3 \times T^4$ solution corresponding to the near horizon limit of a D3-D3 intersection. This is not to be confused with the $\AdS_3 \times S^3 \times T^4$ and $\AdS_3 \times S^3 \times K3$ solutions which describe the D1-D5 system. While the former carries five-form flux, the latter support three-form flux. The two solutions with $T^4$ factors are related by T-duality. Note that the D1-D5 solutions are essentially of Freund-Rubin type. Their Kaluza-Klein spectra are amenable to harmonic analysis and were studied extensively in the early days of AdS/CFT duality \cite{Deger:1998nm,deBoer:1998kjm,Maldacena:1998bw,deBoer:1998us,Aharony:1999ti}.

The calculation of $H_{\tCRc}^{0,0}(S^3 \times T^4)$ is very illustrative. Instead of calculating on $M_7$, we make the transition to the cone
\begin{equation}
  C(S^3) \times T^4 = \bC^2 \times T^4.
\end{equation}
There is an important subtlety here: Since $S^3 \times T^4$ is a direct product, it follows that the contact form $\eta$ dual to the R-symmetry generating vector field $\xi$ has no legs along the $T^4$ factor. Therefore one has to consider the cone over $S^3$ instead of the product $S^3 \times T^4$. For the cohomology groups this means that $H_{\tCRc}^{0,0}$ is isomorphic to $H_{\bar{\partial}}^{0,0}(\bC^2)$ since there are no holomorphic functions on the compact $T^4$. A holomorphic function of fixed $R$-charge is simply a homogeneous polynomial with the $R$-charge being proportional to its degree.

While we emphasized that the geometry in question is different from that of the D1-D5 intersection, it is still interesting to compare this result. In the case of the D1-D5 intersection one argues generally that it is sufficient to consider only the Kaluza-Klein spectrum of six-dimensional supergravity on $\AdS_3 \times S^3$ since the volume of $T^4$ or $K3$ scales in such a way to render fluctuations there redundant \cite{Deger:1998nm,deBoer:1998kjm,Maldacena:1998bw,deBoer:1998us,Aharony:1999ti}. In the case at hand, we find that modes saturating the bound \eqref{eq:unitarity-bound_axio-dilaton} are independent of $T^4$ due to holomorphy.

\subsection{Universal Twist for $Y^{p,q}$}
\label{Sec:ExampleUnivFlow}

A particularly interesting class of the $AdS_3$ solutions we are examining in this paper arises from the twisted compactifications of four-dimensional $\mathcal{N}=1$ gauge theories dual to $\AdS_5 \times Y^{p,q}$ on a Riemann surface \cite{Benini:2015bwz}. For simplicity we will focus on the case of the universal twist where the IR R-symmetry is the same as in the UV and the gauge field is turned on only along the R-symmetry bundle. This class of solutions was first described in section 6.1 of \cite{Gauntlett:2006qw}. Before proceeding with the actual example let us first analyse the holomorphicity constraint $\tCRc Y_0 = 0$.

\subsubsection{The Holomorphicity Constraint}
\label{subSec:HolomCon}
As we described in the previous section the eigenvalues of the operator $\mathcal{L}_0$ saturate the bound $E_0=q$ when the eigenfunctions $Y_0$ of the axio-dilaton fluctuations $B$ are holomorphic with respect to the tangential Cauchy-Riemann operator \eqref{eq:decomposition_exterior_differential}. That is when 
\begin{equation} \label{HolomConstraint}
\tCRc Y_0 = 0 \, .
\end{equation}
One can study $H_{\tCRc}^{0,0}$ using formal methods, yet in certain cases it is also convenient to simply treat the above as a first order PDE and solve it directly. This will allow us to make a connection with the results of \cite{Ardehali:2013xla}. Proceeding in this way, we rewrite equation \eqref{HolomConstraint} as
\begin{equation} \label{ProjCond}
\Pi d Y_0 = 0 \, ,
\end{equation}
where $\Pi$ is a projector $\Pi : T_{\bC} M_7 \to T^{0,1}$. Specifically,
\begin{equation} \label{Projector}
\Pi = \frac{1}{2}\left[ 1+i \mathcal{I} - \eta \otimes \left(\eta\lrcorner\right)\right] \, , 
\end{equation}
with $\mathcal{I}$ being the almost complex structure on $M_6$. We also have
\begin{equation}\label{extderivframes}
dY_0=\hat{e}^\alpha \widehat{E}_\alpha(Y_0) = \hat{e}^\alpha \widehat{E}_\alpha^{\mu} \partial_{\mu}Y_0 \, , \quad \alpha=1,2,...,7 \, ,
\end{equation}
where $\hat{e}^\alpha$ are the vielbein of $\mathcal{M}_7$ with $\hat{e}^7=\eta$ and $\hat{e}^a=e^{-2A} e^a$ for $a=1,...,6$. The $e^a$ are the vielbein of $\mathcal{M}_6$. The scalar mode $Y_0$ decomposes into real and imaginary parts
\begin{equation}
Y_0= Y_r + i Y_{im} \, .
\end{equation}
Using the above (\ref{ProjCond}) reduces to three sets of Cauchy-Riemann equations
\begin{IEEEeqnarray}{rCl}\label{CRframes}
 \widehat{E}_{2j}(Y_r) + \widehat{E}_{2j-1}(Y_{im}) &=& 0 \, , \nonumber\\
 \widehat{E}_{2j-1}(Y_r) - \widehat{E}_{2j}(Y_{im}) &=& 0 \, ,
\end{IEEEeqnarray}
where $j=1, 2, 3$.

\subsubsection{The Wave Functions for $Y^{p,q}$}
\label{subSec:Ypqmodes}
For these backgrounds, the warp factor is trivial. The metric meanwhile is
\begin{equation}
ds_{10}^2=ds^2 (AdS_3) + \frac{3}{4}ds^2_{\mathfrak{g}>1}+\frac{9}{4}ds^2_{\widetilde{Y}^{p,q}} \, ,
\end{equation}
where $\widetilde{Y}^{p,q}$ is the five-dimensional metric of $Y^{p,q}$ fibered over the Riemann surface. Locally, the constant curvature metric over the Riemann surface takes the form
\begin{equation}
ds^2_{\mathfrak{g}>1} = ds^2_{\bH_2} = \frac{1}{x_2^2}\left(dx_1^2 + dx_2^2\right) \, .
\end{equation}
Any Riemann surface of genus $\mathfrak{g}>1$ can be written as quotient of $\bH_2$ with a Fuchsian group $\Gamma$; i.e.~a discrete subgroup of $\SL(2,\bR)$.\footnote{For an introduction, see e.g.~\cite{beardon2012geometry}.} In coordinates that make the Reeb foliation explicit we have
\begin{equation}\label{eq:YpqMetric}
\begin{split}
& ds^2_{\widetilde{Y}^{p,q}}=\frac{1-cy}{6}\left(d\theta^2+\sin^2\theta d\phi^2\right) + \frac{1}{w(y)q(y)}dy^2+\frac{w(y)q(y)}{36}\left(d\beta + c\cos\theta d\phi\right)^2 \\
& + \frac{1}{9}\left(d\psi-\cos\theta d\phi +y\left(d\beta+c\cos\theta d\phi\right) - \frac{dx_1}{x_2}\right)^2 \, ,
\end{split}
\end{equation}
with 
\begin{equation}
w(y) = \frac{2\left(a-y^2\right)}{1-cy} \, , \quad q(y) = \frac{a-3y^2+2cy^3}{a-y^2}\, .
\end{equation}
Comparing with the canonical form of the metric in \cite{Gauntlett:2004yd}, one sees that the fibration is due to the $\frac{dx_1}{x_2}$ term. However, the orbits of the Reeb vector are in general not closed \cite{Gauntlett:2004yd} and thus $\psi$ is not a suitable coordinate to solve the PDE \eqref{HolomConstraint}. After performing the coordinate transformation
\begin{equation} \label{YpqCoordTrans}
\alpha = -\frac{\beta}{6} - c\frac{\psi}{6} \, , \quad \tilde{\psi}=\psi \, ,
\end{equation} 
one has periodic coordinates $\tilde{\psi}$ and $\alpha$ with periods $2\pi$ and $2\pi l$ respectively where
\begin{equation}\label{lDefn}
l = \frac{q}{3q^2 - 2p^2 + p\left(4p^2 - 3q^2\right)^{1/2}} \, .
\end{equation}
In terms of these coordinates, the metric is
\begin{IEEEeqnarray}{rCl}
 ds^2_{\widetilde{Y}^{p,q}} &=& \frac{1-cy}{6}\left(d\theta^2+\sin^2\theta d\phi^2\right) + \frac{1}{w(y)q(y)}dy^2 + \frac{q(y)}{9}\left(d\tilde{\psi}-\cos\theta d\phi -\frac{dx_1}{x_2}\right)^2 \nonumber\\
 &&+ w(y)\left(d\alpha + \frac{ac-2y+y^2c}{6\left(a-y^2\right)}\left(d\tilde{\psi}-\cos\theta d\phi\right)+\frac{2y}{3w(y)}\frac{dx_1}{x_2}\right)^2 \, .
\end{IEEEeqnarray}
Now that the $U(1)$ isometries are explicit, we can make the same ansatz for $Y_0$ as in \cite{Kihara:2005nt}, namely, that $Y_0$ factorizes:
\begin{equation}
Y_0 = e^{\imath N_{\tilde{\psi}} \tilde{\psi} + \imath N_{\phi}\phi + \imath\frac{N_{\alpha}}{l}\alpha}X\left(x_1,x_2\right) \Theta\left(\theta\right)R\left(y\right)\, .
\end{equation}
The $R$-charge $q$ is defined as
\begin{equation}\label{eq:2d_R-charge_defined}
2\partial_{\psi}Y_0 = \imath q Y_0 = \imath\left(2N_{\tilde{\psi}}-\frac{N_{\alpha}}{3l}\right)Y_0 \, .
\end{equation}
Upon combining the real and imaginary parts of $Y_0$, the Cauchy-Riemann equations (\ref{CRframes}) give us three equations, one for the Riemann surface, one for the two-sphere and one for the directions $y$, $\alpha$, which respectively are
\begin{subequations}
  \begin{IEEEeqnarray}{rCl}
    x_2 \left(\partial_{x_1}Y_0 +i\partial_{x_2}Y_0 \right) + \partial_{\psi}Y_0 &=& 0 \, , \\
    \frac{1}{\sin\theta}\partial_{\phi}Y_0 -i\partial_{\theta}Y_0  + \cot\theta\partial_{\tilde{\psi}}Y_0 &=& 0\, , \\
    \frac{A(y)^2}{6G(y)}\partial_{\alpha}Y_0  + \frac{A(y)}{G(y)}y\partial_{\tilde{\psi}}Y_0 +\frac{i}{3}\partial_y Y_0 &=& 0\, . \label{eq:Ypq_Riemann_surface_PDE}
  \end{IEEEeqnarray}
\end{subequations}
Here, we defined
\begin{equation}
A(y)=1-cy \, , \quad G(y)=a+2cy^3-3y^2 \, .
\end{equation}
We solve \eqref{eq:Ypq_Riemann_surface_PDE} in the upper half plane. On $\bH_2$ the general solution is
\begin{equation}\label{eq:Ypq_solution_on_H2}
X(s,\bar{s})=f(s)x_2^{-q/2}\, ,
\end{equation}
where $s=x_1 + \imath x_2$ and $f(s)$ is an arbitrary holomorphic function. The remaining equations can be solved as in \cite{Ardehali:2013xla}. For the fluctuations along the $S^2$,
\begin{equation}
\begin{split}
\Theta (\theta ) & =\frac{\left(\sin\theta\right)^{N_{\phi}+N_{\tilde{\psi}}}}{\left(1+\cos\theta\right)^{N_{\phi}}} 
 = \left(\sin\frac{\theta}{2}\right)^{-N_{\phi}-N_{\tilde{\psi}}}\left(\cos\frac{\theta}{2}\right)^{N_{\phi}-N_{\tilde{\psi}}} P_0^{\left(-N_{\phi}-N_{\tilde{\psi}} , N_{\phi}-N_{\tilde{\psi}}  \right)} (\cos\theta) \, .
\end{split}
\end{equation}
In the second step we rewrote the result in terms of Jacobi polynomials. Finally for the fluctuations along $y, \alpha$ we get,
\begin{equation}
R\left(y\right) = \prod_{i=1}^3\left(y-y_i\right)^{a_i} ,
\end{equation}
where $y_i$ are the roots of the polynomial $G(y)=0$ and 
\begin{equation}
a_i = \frac{N_{\alpha}}{12l}\left(\frac{1}{y_i} -1\right) + \frac{N_{\tilde{\psi}}}{2}\, ,\quad  i=1, 2, 3 \, .
\end{equation}

On $\bH_2$, the situation is thus clear: There is an infinite number of solutions to \eqref{eq:Ypq_solution_on_H2}. For each of these there is a copy of the corresponding solutions of the scalar Laplacian on $Y^{p,q}$ at the unitarity bound as in \cite{Ardehali:2013xla} and thus a copy of a subsector of the mesonic chiral ring of the four-dimensional $\mathcal{N}=1$ theory.

The crucial question is of course which of these modes survive the transition to the quotient $\bH_2 / \Gamma$.\footnote{We would like to thank the referee at JHEP for observations that led to the following discussion.}
Instead of focussing on a specific choice of $\Gamma$, let us first consider a generic $\SL(2,\bR)$ transformation which acts on $s \in \bH_2$ as
\begin{equation}
  s \mapsto \frac{A s + B}{C s + D} \equiv s^\prime, \qquad
  \begin{pmatrix} A & B \\ C & D \end{pmatrix} \in \SL(2, \bR).
\end{equation}
While $ds_{\mathfrak{g}>1}^2$ is invariant under this transformation, the cross term in \eqref{eq:YpqMetric} is not. Instead, one finds that
\begin{equation}
  \frac{dx_1}{x_2} \mapsto \frac{dx_1}{x_2} + 2 d \left( \arctan \frac{C x_2}{C x_1 + D} \right).
\end{equation}
Since the mismatch is exact, it can be absorbed by a compensating transformation
\begin{equation}
  \psi \mapsto \psi + 2 \arctan \frac{C x_2}{C x_1 + D}.
\end{equation}
We return to the wave function $Y_0$. In terms of the coordinates of \eqref{eq:YpqMetric}, we need to study the transformation behavior of $e^{\frac{\imath q}{2} \psi} x_2^{-q/2} f(s)$. Using $\arctan u = \frac{\imath}{2} \log \frac{1 - \imath u}{1 + \imath u}$, we see that the exponential factor transforms as
\begin{equation}
  e^{\frac{\imath q}{2} \psi} \mapsto e^{\frac{\imath q}{2} \psi} \left( C s + D \right)^{q/2} \left( C \bar{s} + D \right)^{-q/2}.
\end{equation}
Similarly, $x_2^{-q/2} \mapsto x_2^{-q/2} (C s + D)^{q/2} (C \bar{s} + D)^{q/2}$. Thus we arrive at the conclusion that invariance of the wavefunction under $\Gamma$ requires the function $f(s)$ to transform as
\begin{equation}\label{eq:modular_form}
  f(s) \mapsto (Cs + D)^{-q} f(s).
\end{equation}
One recognizes the transformation behavior of a modular form of weight $-q$. With $f$ being a modular form on the Riemann surface $\bH_2 / \Gamma$, \eqref{eq:modular_form} holds $\forall \left( \begin{smallmatrix} A & B \\ C & D \end{smallmatrix}\right) \in \Gamma$, yet not for generic elements of $\SL(2,\bR)$. At this point one should wonder about the sign of $q$. The interpretation of $f$ as a modular form of weight $-q$ suggests $q \leq 0$. Moreover, the wave function $Y_0$ is singular at $x_2 = 0$ for positive values of $q$. On the other hand, the definition of the 2D R-charge in \eqref{eq:2d_R-charge_defined} is identical to that of the 4D R-charge, which leads to $q \geq 0$ which is consistent with our conventions in section \ref{sec:minimally_coupled_scalar}. Whatever the resolution to this question, the above exhibits an injective map from the set of modular forms of weight $-q$ on the Riemann surface to the set of short multiplets of the CFT.

\section{The Dilatino and Superconformal Multiplets}
\label{sec:dilatino_analysis}

Our discussion of the axio-dilaton and spin 2 fluctuation in section \ref{sec:minimally_coupled_scalar} relied heavily on holomorphy. Our discussion in the introduction emphasized however that holomorphy and its relation to the unitarity bound is just one tool that one can exploit in the Kaluza-Klein analysis. In this section we will instead focus on superconformal symmetry using the dilatino as an example. The situation is simplified by the fact that its fluctuations also decouple from the rest of the spectrum -- a fact that also holds for fluctuations of the three-form that we will turn to in section \ref{sec:three-form}. We will be able to prove that any wave function discussed in section \ref{sec:minimally_coupled_scalar} -- not just those satisfying the bound -- immediately defines wave functions in the dilatino spectrum. The corresponding modes have the correct quantum numbers to be superpartners of the axio-dilation fluctuations, which agrees with the form of the supersymmetry transformations of type IIB supergravity. Note however that we do not calculate the complete dilatino spectrum which should also contain modes that lie in other multiplets.

\subsection{Some Lessons from the Superconformal Algebra}
\label{sec:superconformal_algebra}

In order to get some intuition, we will review some elementary aspects of the representation theory of the $\mathcal{N}=2$ algebra. For details see \cite{Bars:1982ep,Gunaydin:1986fe,Lerche:1989uy} and references therein. What is relevant for our analysis is the $\osp (2 \vert 2)$ subalgebra
\begin{IEEEeqnarray}{rClCrCl}\label{eq:N2_algebra}
  \lbrack \bar{L}_0, \bar{L}_{\pm 1} \rbrack &=& \mp \bar{L}_{\pm 1}, &\qquad&
  \lbrack J_0, \bar{L}_{\pm 1} \rbrack &=& 0, \nonumber\\
  \lbrack \bar{L}_0, G^+_{\pm 1/2} \rbrack &=& \mp \frac{1}{2} G^+_{\pm 1/2}, &\qquad&
  \lbrack J_0, G^+_{\pm 1/2} \rbrack &=& G^+_{\pm 1/2}, \nonumber\\
  \lbrack \bar{L}_0, G^-_{\pm 1/2} \rbrack &=& \mp \frac{1}{2} G^-_{\pm 1/2}, &\qquad&
  \lbrack J_0, G^-_{\pm 1/2} \rbrack &=& - G^-_{\pm 1/2}.
\end{IEEEeqnarray}

Now, as we saw in section \ref{sec:minimally_coupled_scalar}, any element of the Kohn-Rossi cohomology group $H_{\tCRc}^{0,0}$ defines a scalar operator
\begin{equation}
  \ket{\frac{q+2}{2}} \otimes \ket{\frac{q+2}{2}, q}
\end{equation}
that does not saturate the unitarity bound yet is conjectured to lie in a short multiplet. Direct application of the algebra shows that there are two candidates for the chiral primary:
\begin{IEEEeqnarray}{rCl}
  G^+_{1/2} \ket{\frac{q+2}{2}, q} &=& \ket{\frac{q+1}{2}, q + 1}, \qquad
  \bar{L}_1 \ket{\frac{q+2}{2}, q} = \ket{\frac{q}{2}, q}.
\end{IEEEeqnarray}
Since the original state is bosonic and there are no fermionic operators in the left handed algebra, the former possibility would imply that the chiral primary is fermionic, while the latter case leads to a bosonic state. In both cases we cannot make a definite statement about the overall spin of the state we are looking for since it is possible that we would have to act with some power of $L_1$ as well. It is thus conceivable that the chiral primary in question is actually the spin 2 fluctuation we found previously. However, one should keep in mind that in the four-dimensional theories the axio-dilaton and spin 2 fluctuations lie in different multiplets \cite{Eager:2012hx}.

\subsection{The Dilatino}
\label{subsec:theDilatino}

Schematically, the supersymmetry variation of the dilaton is $\delta \Phi \sim \bar{\epsilon} \lambda$ and similar for the axion. It follows that some of the fluctuations of the dilatino and all of the fluctuations of the axio-dilaton should be related by the action of one of the supercharges $\bI \otimes G^\pm_{\pm 1/2}$. With this in mind we consider fluctuations of the dilatino.

\subsubsection{The Supersymmetry Spinor}

To begin, we need to recall some properties of the background supersymmetry spinor. Our discussion mainly follows \cite{Kim:2005ez}. However, see also appendix A of \cite{Donos:2008ug}.

The supersymmetry variation of the gravitino imposes that the ten-dimensional Killing spinor satisfies\footnote{Since we are working in flat indices, we emphasize that there is a difference between $\hat{F}_{A_1 \dots A_5} = \hat{E}_{A_1}^{M_1} \dots \hat{E}_{A_5}^{M_5} F_{M_1 \dots M_5}$ and $F_{A_1 \dots A_5} = E_{A_1}^{M_1} \dots E_{A_5}^{M_5} F_{M_1 \dots M_5}$.
}
\begin{equation}
  0 = \hat{\nabla}_A \epsilon + \frac{\imath}{480} \hat{F}_{A_1 \dots A_5} \Gamma^{A_1 \dots A_5} \Gamma_A \epsilon.
\end{equation}
Comparing the spin connections on the warped and un-warped frames on $M_{10}$ ($\hat{e}^{A_1} = e^{A} e^{A_1}$) yields
\begin{equation}
  \hat{\omega}^{AB} = \omega^{AB} + 2 \hat{e}^{[A} \hat{E}^{B]} (A).
\end{equation}
Writing the supersymmetry spinor as $\epsilon = \left( \begin{smallmatrix} 1 \\ 0 \end{smallmatrix} \right) \otimes \epsilon \otimes \zeta$, it follows that
\begin{IEEEeqnarray}{rCl}\label{eq:KSE_zeta}
  0 &=& \left( \slashed{\partial} A - \imath n + \frac{1}{2} e^{-4A} F^{\beta\gamma} \gamma_{\beta\gamma} \right) \zeta, \nonumber\\
  0 &=& \left( \nabla_\alpha + \frac{1}{2} \gamma_{\alpha\beta} \partial^\beta A - \frac{1}{4} e^{-4A} F_{\beta\gamma} \gamma^{\beta\gamma} \gamma_\alpha \right) \zeta.
\end{IEEEeqnarray}
This implies that $\zeta^c = C_7 \zeta^*$ satisfies
\begin{IEEEeqnarray}{rCl}
  0 &=& \left( \slashed{\partial} A - \imath n - \frac{1}{2} e^{-4A} F^{\beta\gamma} \gamma_{\beta\gamma} \right) \zeta^c, \nonumber\\
  0 &=& \left( \nabla_\alpha + \frac{1}{2} \gamma_{\alpha\beta} \partial^\beta A + \frac{1}{4} e^{-4A} F_{\beta\gamma} \gamma^{\beta\gamma} \gamma_\alpha \right) \zeta^c.
\end{IEEEeqnarray}

In principle one wants to calculate $\pounds_\xi \zeta$ to confirm that the $R$-charge of $\zeta$ is $\imath n \zeta$ and that of the conjugate spinor $- \imath n$. Instead we just note that the three-form $\Omega = \zeta^T \gamma_{(3)} \zeta$ in \cite{Kim:2005ez} satisfies $\pounds_\xi \Omega = 2 \imath n \Omega$. This implies that both spinors have the correct $R$-charge.

\subsubsection{Dilatino Fluctuations}

The equation of motion for fluctuations of the dilatino is (see e.g.~\cite{Schwarz:1983qr,DeWolfe:2002nn,Argurio:2006my})
\begin{equation}
  \Gamma^A \hat{\nabla}_A \lambda = \frac{\imath}{240} \hat{F}_{A_1 \dots A_5} \Gamma^{A_1 \dots A_5} \lambda.
\end{equation}
The dilatino is chiral, $\Gamma^{(10)} \lambda = - \sigma_3 \lambda = \lambda$. Thus $\lambda = \left( \begin{smallmatrix} 0 \\ 1 \end{smallmatrix} \right) \otimes \lambda \otimes \chi$ and
\begin{IEEEeqnarray}{rCl}
  0 &=& \slashed{\nabla} \lambda \otimes \chi - \imath \lambda \otimes \left( \slashed{\nabla} \chi + \frac{9}{2} \slashed{\partial} A \chi + \frac{1}{2} e^{-4A} \slashed{F} \chi \right).
\end{IEEEeqnarray}
Therefore the mass operator for dilatino fluctuations is
\begin{equation}
  \mathcal{L}_{1/2} \equiv \slashed{\nabla} + \frac{9}{2} \slashed{\partial} A + \frac{1}{2} e^{-4A} \slashed{F}.
\end{equation}

Some experimentation along the lines of \cite{Pope:1982ad} suggests that given an eigenfunction $\tilde{Y}$ of the operator $\mathcal{L}_0$ one can construct eigenmodes of $\mathcal{L}_{1/2}$ by considering $\tilde{Y} \zeta^c$ and $\slashed{\partial} \tilde{Y} \zeta^c$.\footnote{The ansatz used here is also indebted to a series of discussions with Y.~Tachikawa concerning the equivalent problem in the Sasaki-Einstein case.} Following this line of thinking one finds
\begin{IEEEeqnarray}{rCl}
  \mathcal{L}_{1/2} \tilde{Y} \zeta^c &=& \slashed{\partial} \tilde{Y} \zeta^c + \tilde{Y} \slashed{\partial} A \zeta^c + \frac{\imath n}{2} \tilde{Y} \zeta^c, \nonumber\\
  \mathcal{L}_{1/2} (\slashed{\partial} \tilde{Y} \zeta^c) &=& \left[ - \mathcal{L} \tilde{Y} - \frac{1}{2} \partial^a \tilde{Y} \partial_a A - \frac{5}{2} \partial_a \tilde{Y} \partial_b A \gamma^{ab} + \frac{3}{4} e^{-4A} \partial_a \tilde{Y} F_{bc} \gamma^{abc} + \frac{3}{2} e^{-4A} \partial_a \tilde{Y} F^{ab} \gamma_b \right] \zeta^c \nonumber\\
  &=& - \mathcal{L} \tilde{Y} \zeta^c + \partial^a \tilde{Y} \partial_a A \zeta^c - \partial_a \tilde{Y} \partial_b A \gamma^{ab} \zeta^c - \frac{3 \imath n}{2} \slashed{\partial} \tilde{Y} \zeta^c.
\end{IEEEeqnarray}
Note that in going from the second line to the third we made use of the algebraic equation for $\zeta^c$. The choice $\tilde{Y} = e^{- A} Y_0$ turns out to lead to a diagonalizable system:
\begin{equation}
  \mathcal{L}_{1/2}
  \begin{pmatrix} e^{-A} Y_0 \zeta^c \\ e^{-A} \slashed{\partial} Y_0 \zeta^c \end{pmatrix}
  =
  \begin{pmatrix} \frac{\imath n}{2} & 1 \\ - \mathcal{L}_0 & - \frac{3\imath n}{2} \end{pmatrix}
  \begin{pmatrix} e^{-A} Y_0 \zeta^c \\ e^{-A} \slashed{\partial} Y_0 \zeta^c \end{pmatrix}.
\end{equation}
The eigenvalues of the mass matrix are $\imath \left( \pm \sqrt{\mathcal{L}_0 + 1} - \frac{n}{2} \right)$.\footnote{\label{fn:dilatino_eigenvectors} For reference, the eigenvectors are
  \begin{IEEEeqnarray*}{rl}
    &\imath e^{-A} Y_0 \zeta^c (\sqrt{\mathcal{L}_0 + 1} + n) - e^{-A} \slashed{\partial} Y_0 \zeta^c (\mathcal{L}_0), \nonumber\\
    &\imath e^{-A} Y_0 \zeta^c (\sqrt{\mathcal{L}_0 + 1} - n) + e^{-A} \slashed{\partial} Y_0 \zeta^c (\mathcal{L}_0).
  \end{IEEEeqnarray*}
  Clearly, contracting with $(\zeta^c)^\dagger$ yields $Y_0$. This shows that one should be able to map some of the eigenmodes of $\mathcal{L}_{1/2}$ to the set of eigenmodes of $\mathcal{L}_0$ by simply contracting with $(\zeta^c)^\dagger$. This is of course just the inverse of the supersymmetry transformation that mapped axio-dilaton fluctuations to dilatino ones.
}
Setting $n = 1$ and labelling the corresponding masses as $m_\pm$, we have
\begin{equation}
  m_\pm = E_0 + 1 \pm \frac{1}{2}.
\end{equation}
Now, recall that $h - \bar{h} = s = \pm 1/2$ and that $h + \bar{h} = \Delta = \vert m \vert + 1$. It follows that we have a mode with $s = 1/2$, $2\bar{h} = E_0 + 1$ and $R$-charge $q - 1$ and another mode with $s = -1/2$, $2\bar{h} = E_0 + 3$ and $R$-charge $q + 1$. At the unitarity bound $E_0 = q$ these are not chiral primaries. However, they are superpartners of the axio-dilaton mode, as we verify that the dilatino states correspond to the action of $\bI \otimes G^-_{\pm 1/2}$ on the axio-dilaton state.

In the above discussion the sign of the helicity $s$ followed from consistency. The ``wrong'' choice of sign leads to modes with $2\bar{h} = q + 2$ which is not possible for a mode which lies in the same multiplet as the mode of section \ref{sec:minimally_coupled_scalar}, yet has opposite spin statistics.

One might wonder whether the spinor $\zeta$ -- instead of its conjugate -- might lead to additional eigenmodes of the dilatino. An identical calculation to the above yields
\begin{IEEEeqnarray}{rCl}
  \mathcal{L}_{1/2} (\slashed{\partial} \tilde{Y} \zeta) &=& \left[ - \mathcal{L} \tilde{Y} - \frac{1}{2} \partial_a \tilde{Y} \partial^a A - \frac{5}{2} \partial_a \tilde{Y} \partial_b A \gamma^{ab} + \frac{1}{4} e^{-4A} \partial_a \tilde{Y} F_{bc} \gamma^{abc} - \frac{7}{2} e^{-4A} \partial_a \tilde{Y} F^{ab} \gamma_b  \right] \zeta, \nonumber\\
  \mathcal{L}_{1/2} \tilde{Y} \zeta &=& \slashed{\partial} \tilde{Y} \zeta - \tilde{Y} \slashed{\partial} A \zeta + \frac{5\imath n}{2} \tilde{Y} \zeta.
\end{IEEEeqnarray}
The point is that it seems impossible to use the algebraic identity for $\zeta$ to further simplify the first of these since the terms of $\mathcal{O}(F)$ differ by a factor of $- 14$ while acting with $\gamma_a$ on the algebraic equation in \eqref{eq:KSE_zeta} yields a relative factor of $2$.

\section{The Three Form and Betti Multiplets}
\label{sec:three-form}

In the final part of this paper we will extend the methods used in section \ref{sec:minimally_coupled_scalar} to study fluctuations of the three-form. For simplicity, we restrict to $n = +1$.

\subsection{Deformed Laplace Operators}
\label{sec:deformed_laplace_operators}

As alluded in section \ref{sec:minimally_coupled_scalar}, the operator $\mathcal{L}_0$ can in fact be regarded as a deformation of the usual Laplace operator. In order to make this relation clear we need to review some aspects of the Hodge dual and the resulting inner product on $p$-forms.

Consider a $d$-dimensional (compact) manifold of signature $t$ and $\alpha \in \Omega^k$, $\beta \in \Omega^l$. The Hodge star is defined by
\begin{IEEEeqnarray}{rCl}\label{eq:Hodge_star_defined}
  \star \bar{\alpha} \wedge \beta &=& \frac{1}{k!} \bar{\alpha}^{m_1 \dots m_k} \beta_{m_1 \dots m_k} \vol \equiv \langle \alpha, \beta \rangle \vol.
\end{IEEEeqnarray}
Occasionally we will add a subscript to denote the metric used to define the Hodge star. E.g.~for $ds_{10A}^2 = e^{2A} ds_{10}^2$, $\star_{10A}$ is the Hodge dual induced by the warped ten-dimensional metric, $\star_{10}$ its unwarped cousin. Appendix \ref{sec:appendix_hodge} contains a number of technical results that we will use extensively.

In order to define a Laplace operator one considers the canonical inner product on $\Omega^k$:
\begin{equation}
  ( \alpha, \beta ) \equiv \int \star \bar{\alpha} \wedge \beta.
\end{equation}
This product leads to the notion of adjoint operators such as $(\alpha, d^*\beta) = (d \alpha, \beta)$ which in turn allow the definition of the de Rham Laplacian via $\Delta = \{ d, d^* \}$. In our conventions one finds
\begin{equation}
  d^* \alpha = (-1)^{kd + t} \star d \star \alpha.
\end{equation}

However, the inner product $(\cdot, \cdot)$ is not unique and so neither is $d^*$. Indeed, the crucial if simple insight in section \ref{sec:minimally_coupled_scalar} was to normalize the inner product on scalars with a factor of $e^{8A}$. Generalizing this we introduce the deformed inner product
\begin{equation}\label{eq:deformed_inner_product}
  ( \alpha, \beta )_c \equiv \int e^{cA} \star \bar{\alpha} \wedge \beta
\end{equation}
where we have introduced the constant $c \in \bR$ and tacitly assumed that the warp-factor is sufficiently well behaved for the integral to converge. The logical next step is to consider deformed adjoints and Laplace operators in terms of the deformed inner product \eqref{eq:deformed_inner_product}. For
\begin{IEEEeqnarray}{rCl}\label{eq:twisted_adjoints}
  (d + c dA \wedge)^*_c &\equiv& (-1)^{kd+t} \star d \star = d^*, \nonumber\\
  d^*_c &\equiv& (-1)^{kd+t} \star d \star - c dA \lrcorner = d^* - c dA \lrcorner
\end{IEEEeqnarray}
one verifies that $(d\alpha, \beta)_c = (\alpha, d_c^* \beta)$. Instead of introducing further symbols we will denote the deformed de Rham Laplacian by
\begin{equation}
  \{ d, d_c^* \} = d d_c^* + d_c^* d.
\end{equation}
For scalar functions on $M_7$ one verifies that
\begin{equation}
  \mathcal{L}_0 = \left. \{ d, d_8^* \} \right\vert_{\Omega^0}.
\end{equation}

In the above discussion we defined $d_c^*$ as the adjoint of the exterior derivative with respect to the deformed inner product $(\cdot, \cdot)_c$. However, we could have just as well defined $d_c^*$ in terms of the inner product $(\cdot, \cdot)$ after rescaling the metric on $M_7$ by a suitable power of $e^A$, with the weight of the exponential depending on the degree of the form. It follows that the usual theorems that are familiar from de Rham and Dolbeault cohomology apply -- most notably Hodge decomposition and the existence of a complete set of orthogonal eigenfunctions of the Laplace operator.

\subsection{Gauge Fixing of the Three-Form Equations}
\label{sec:three-form_gauge_fixing}

Linearising the equation of motion for the three-form, $d_{10A}^* G = \imath G \lrcorner_{10A} F_{5}$, leads to ($G = da$)
\begin{equation}
  d_{10A}^* d a = \imath da \lrcorner_{10A} F_5 = \imath e^{-6A} da \lrcorner_{10} F_5.
\end{equation}
Using a standard decomposition
\begin{equation}
  a_2(M_{10}) = \sum a_k(\AdS_3) \otimes Y_{2-k}(M_7)
\end{equation}
as well as \eqref{eq:F5} one arrives at the set of equations\footnote{This uses
\begin{IEEEeqnarray*}{rl}
  &e^{2A} d_{10A}^* d (a_k Y_{2-k}) = \nonumber\\
  &= d^* d a_k Y_{2-k} + a_k d^* d Y_{2-k} - da_k \star d \star Y_{2-k} - \star d \star a_k dY_{2-k} + 4 (-1)^k da_k dA \lrcorner Y_{2-k} - 4 a_k dA \lrcorner dY_{2-k}.
\end{IEEEeqnarray*}
}
\begin{subequations}\label{eq:three-form_equations_not_split}
  \begin{IEEEeqnarray}{rCl}
    0 &=& d^* d a_2 Y_0 + a_2 (d^* d Y_0 - 4 dA \lrcorner dY_0) - da_1 (\star d \star Y_1 + 4 dA \lrcorner Y_1) \nonumber\\
    &&+ \imath e^{-4A} ( \star da_0 Y_2 \lrcorner F - \star a_1 dY_1 \lrcorner F ), \\
    0 &=& d^* d a_1 Y_1 + a_1 (d^*d Y_1 - 4 dA \lrcorner dY_1) - da_0 (\star d \star Y_2 - 4 dA \lrcorner Y_2) - \star d \star a_2 dY_0 \nonumber\\
    &&+ \imath e^{-4A} ( \star da_1 Y_1 \lrcorner F + \star a_2 dY_0 \lrcorner F), \\
    0 &=& d^* d a_0 Y_2 + a_0 (d^* d Y_2 - 4 dA \lrcorner dY_2) - \star d \star a_1 dY_1 \nonumber\\
    &&- \imath e^{-4A} \left[ \star da_2 Y_0 F + a_0 dY_2 \lrcorner \left( \frac{J \wedge \rho \wedge \eta}{8} - e^{4A} \star (dA \wedge \eta) \right) \right].
  \end{IEEEeqnarray}
\end{subequations}
Each equation contains terms of the form $d^*d Y_k - 4 dA \lrcorner dY_k$ as well as $d^* Y_k - 4 dA \lrcorner Y_k$. If we impose the gauge condition
\begin{equation}
   d_4^* Y_k = 0,
\end{equation} 
the latter vanish, while the former become deformed Laplacians:
\begin{equation}
  d^* d Y_k - 4 dA \lrcorner dY_k + d (d^* Y_k - 4 dA \lrcorner Y_k) = \{ d, d_4^* \} Y_k.
\end{equation}
In other words, the twisted adjoints and Laplace operators defined in the previous section appear to be a good language to describe the equations of motion.

One might wonder whether this gauge condition is consistent. Continuing from the discussion at the end of section \ref{sec:deformed_laplace_operators}, we assume that Hodge decomposition holds. Then we can decompose any form $Y_k$ into a closed, co-closed and harmonic part:
\begin{equation}
  Y_k = d y_{k-1} + d_4^* y_{k+1} + y_k, \qquad  \{ d, d_4^* \} y_k = 0.
\end{equation}
By a gauge transformation, we can set $y_{k-1}$ to zero and since harmonic forms are closed and co-closed it follows that $d_4^* Y_k = 0$.

To proceed, we assume wave functions $Y_k$ of different degree $k$ to be orthogonal and similarly for the modes $a_k$. Then the above decompose into three equations for $a_2$,
\begin{subequations}\label{eq:three_form_equations_a2}
  \begin{IEEEeqnarray}{rCl}
    0 &=& \star da_2 \otimes Y_0 F, \\
    0 &=& d^* d a_2 \otimes Y_0 + a_2 \otimes \{d, d_4^*\} Y_0, \\
    0 &=& \star d \star a_2 \otimes dY_0 - \imath e^{-4A} \star a_2 \otimes dY_0 \lrcorner F,
  \end{IEEEeqnarray}
\end{subequations}
three equations for $a_1$
\begin{subequations}\label{eq:three_form_equations_a1}
  \begin{IEEEeqnarray}{rCl}
    0 &=& \star d \star a_1 \otimes dY_1, \\
    0 &=& - da_1 \otimes d_4^* Y_1 + \imath e^{-4A} \star a_1 \otimes dY_1 \lrcorner F, \\
    0 &=& d^* d a_1 \otimes Y_1 + a_1 \otimes \{d, d_4^*\} Y_1 + \imath e^{-4A} \star da_1 \otimes Y_1 \lrcorner F,
  \end{IEEEeqnarray}
\end{subequations}
and three equations for $a_0$
\begin{subequations}\label{eq:three_form_equations_a0}
  \begin{IEEEeqnarray}{rCl}
    0 &=& \star da_0 \otimes Y_2 \lrcorner F, \\
    0 &=& da_0 \otimes d_4^* Y_2, \\
    0 &=& d^* d a_0 \otimes Y_2 
    + a_0 \otimes \left[ \{d, d_4^* \} Y_2 - \imath e^{-4A} dY_2 \lrcorner \left( \frac{J \wedge \rho \wedge \eta}{8} - e^{4A} \star (dA \wedge \eta) \right) \right].
  \end{IEEEeqnarray}
\end{subequations}

\subsection{Betti Multiplets}
\label{sec:betti_multiplets}

Equations \eqref{eq:three_form_equations_a2}, \eqref{eq:three_form_equations_a1} and \eqref{eq:three_form_equations_a0}, although quite complicated, simplify considerably if we assume the wave functions to be closed, $d Y_k = 0$. Since we assumed them to be also orthogonal, it is better to also assume that they are not exact. Due to the gauge condition they are then harmonic with respect to the deformed Laplacian, $\{d, d_4^*\} Y_k = 0$. Finally, we observe that the R-charge vanishes if they are horizontal, since $\pounds_\xi Y_k = d (\iota_\xi Y_k)$.

\begin{itemize}
  \item From equations \eqref{eq:three_form_equations_a0} it follows that for every element of $H^2(M_7)$ that is orthogonal to $F$ ($Y_2 \lrcorner F = 0$) there is a massless scalar $a_0$. From the AdS/CFT dictionary it follows that $\Delta_\pm = 1 \pm 1$. If moreover $Y_2$ is horizontal, the $R$-charge is zero.

  \item The situation is a little more complicated for the one-form $a_1$. For $dY_1 = 0$, the first equation in \eqref{eq:three_form_equations_a1} no longer imposes the constraint $d^* a_1 = 0$ while the remaining ones are gauge-invariant under $a_1 \mapsto a_1 + d\lambda$. This behaviour is actually familiar from the Freund-Rubin case. See e.g.~the discussion in \cite{Duff:1986hr}. The second equation reduces to the constraint $dY_1 \lrcorner F = 0$. To deal with the final equation, we take the ``square root'' of the Laplacian by defining $Q = \imath \star_3 d$. The equation is now
  \begin{equation}
    0 = Q^2 a_1 \otimes Y_1 + Q a_1 \otimes e^{-4A} Y_1 \lrcorner F.
  \end{equation}
  If the two terms are linearly independent, we have $Q a_1 = 0$ and thus $da_1 = 0$. However, since the universal cover of anti-de Sitter space has trivial fundamental group this means that $a_1$ is pure gauge. Thus, the existence of a non-trivial solution requires the existence of some constant $y_1 \in \bC$ such that $Y_1 \lrcorner F = \imath e^{4A} y_1 Y_1$. Together with the gauge condition this implies that the constraint is satisfied as
  \begin{equation}
    dY_1 \lrcorner F = d^* (Y_1 \lrcorner F) = \imath y_1 d^* (e^{4A} Y_1) = \imath y_1 e^{4A} (d^* Y_1 - 4 dA \lrcorner Y_1) = 0.
  \end{equation}
  The fluctuation equation reduces to
  \begin{equation}
    0 = (Q^2 + \imath y_1 Q) a_1,
  \end{equation}
  with eigenvalues $Q = 0$ and $Q = - \imath y_1$. Again, the $Q = 0$ eigenvalue leads to a $a_1$ being pure gauge. The mass is given by $Q^2$ and thus $y_1^2$. The mass-dimension formula for a one-form in $\AdS_3$ is $\Delta_\pm = 1 \pm \vert m \vert$. Since these modes have spin $1$ we have $h - \bar{h} = \pm 1$ and thus $2 \bar{h} = 1 \mp 1 \pm \vert m \vert$. In appendix \ref{sec:product_of_kahler_einstein} we calculate $y_1$ for some simple examples.

  \item The simplest equations are those determining the two-form $a_2$. The first of \eqref{eq:three_form_equations_a2} imposes that $a_2$ is closed. However, $H^2(\AdS_3) = 0$ and one sees immediately that no such fluctuations exist.
\end{itemize}
This concludes our discussion of the three-form equations.

\section{Future Directions}

Our results point to a number of interesting directions for future research. To begin there is the clarification of the example considered in section \ref{Sec:ExampleUnivFlow} with regards to the quotient $\bH_2 / \Gamma$. Subsequently generalizing the analysis done there to the large number of solutions present in \cite{Benini:2015bwz} will give an answer to the question how the Hilbert spaces of the two-dimensional theories arise from those of their four-dimensional avatars.

As we mentioned in the introduction, one of the most interesting discoveries of \cite{Benini:2015bwz} is the mixing of the UV $R$- and and baryonic-symmetries. Being mesonic operators however, the supergravity fluctuations we discuss here are not sensitive to this effect; an interpretation that is consistent with both our discussion of the unitarity bound and our results concerning the superpartners in the dilatino spectrum. Baryonic operators dual to wrapped branes on the other hand are sensitive to this effect. This suggests that one should perform a careful analysis of these \cite{Gubser:1998fp,Berenstein:2002ke}.

While we used the approach of \cite{Ardehali:2013xla} to calculate $H_{\tCRc}^{0,0}$ directly, one should not forget that a large number of results in four dimensions have been obtained by considering the Calabi-Yau cone instead of its Sasaki-Einstein base. The situation is more complicated in the case at hand since $M_7$ is constrained by equation \eqref{eq:M6_curvature_constraint} and the cones $C(M_7)$ are not K\"ahler, yet the eight-dimensional perspective should still be an interesting avenue to explore.

Turning to questions in supergravity and Kaluza-Klein theory, one would like to complete the analysis of the spectrum started here. The most interesting question here might be whether higher cohomology groups $H_{\tCRc}^{p,q}$ contribute to the chiral ring beyond $H_{\tCRc}^{0,0}$. Comparing sections \ref{sec:minimally_coupled_scalar} and \ref{sec:three-form}, one might thus wonder whether it is possible to extract further information from the three-form equations by using holomorphy as a guiding principle. We have actually attempted to do so following \cite{Schmude:2013dua}, yet were not able to find any further modes with wave functions satisfying $\tCRc Y_k = 0$. While this might indicate that there are no such modes in the spectrum of the three-form, one should study the diagonalization of the system \eqref{eq:three-form_equations_not_split} more carefully. Once one has achieved a sufficient understanding of the Kaluza-Klein spectrum, one should be able to calculate the elliptic genus as in \cite{deBoer:1998kjm} to obtain results similar to those of \cite{Eager:2012hx,Eager:2013mua}; i.e.~as a weighted sum over $H_{\tCRc}^{p,q}$. Finally, one can of course generalize our approach to backgrounds including more general fluxes such as \cite{Donos:2008ug} or in different dimensions. We hope to return to these topics in the future.

\section*{Acknowledgements}

We would like to thank Nikolay Bobev, Eoin \'O Colg\'ain, Carlos Hoyos, Dario Martelli, Patrick Meessen, Diego Rodriguez-Gomez, Andres Vi\~na, Simon Wood and Konstantin Zarembo for valuable discussions and correspondence. J.S.~would like to thank the organizers of the workshop ``Holography and Dualities 2016: New Advances in String and Gauge Theory'' at Nordita for hospitality during the later stages of this project. The work of J.S.~is funded by an EU Marie Curie--Clar\'in--COFUND fellowship of the Regional Government of Asturias. O.V.~is partially supported by the Ramon y Cajal fellowship
RYC-2012-10370.

\appendix

\section{Conventions and Useful Expressions}

\subsection{Differential Forms and the Hodge Star}
\label{sec:appendix_hodge}

In terms of indices, the definition of the Hodge star in \eqref{eq:Hodge_star_defined} translates to ($\epsilon_{123 \dots d} = 1$)
\begin{equation}\label{eq:Hodge_defined_in_indices}
  \star \alpha_{m_1 \dots m_{d-k}} = \frac{\sqrt{g}}{k!} \epsilon_{m_1 \dots m_{d-k}}^{\phantom{m_1 \dots m_{d-k}} n_1 \dots n_k} \alpha_{n_1 \dots n_k}.
\end{equation}
One verifies
\begin{IEEEeqnarray}{rCl}
  \star \star \alpha &=& (-1)^{k(d-k)+t} \alpha, \nonumber\\
  \star (\beta \wedge \alpha) &=& \frac{\sqrt{g}}{k! l!} \epsilon_{m_1 \dots m_{d-(k+l)}}^{\phantom{m_1 \dots m_{d-(k+l)}} n_1 \dots n_{k+l}} \beta_{n_1 \dots n_l} \alpha_{n_{l+1} \dots n_{k+l}} = \frac{1}{l!} \beta^{n_1 \dots n_l} (\star\alpha)_{m_1 \dots m_{d-(k+l)} n_1 \dots n_l} \nonumber\\
  &=& (-1)^{l[d-(k+l)]} \beta \lrcorner \star \alpha,
  \end{IEEEeqnarray}
where
\begin{equation}
   \beta \lrcorner \alpha \equiv \frac{1}{l!} \beta^{n_1 \dots n_l} \alpha_{n_1 \dots n_l m_k-l+1 \dots m_k}
\end{equation}
and we've assumed that $d-k \geq l$. 

For our background \eqref{eq:IIB_metric} the cotangent bundle decomposes as
\begin{equation}
  T^*M_{10} = T^* \AdS_3 \oplus T^* M_7 = T^*\AdS_3 \oplus T^*M_6 \oplus \bR \eta.
\end{equation}
We need to consider how the various Hodge star operators are related. To do so, we need to recall how the Hodge star decomposes in the generic case. Let $(V, \langle, \rangle)$ be a vector space with an inner product of signature $t$. Moreover, there is a decomposition $V = W_1 \oplus W_2$ compatible with the inner product. Assume that the signature of $\langle, \rangle_1$ is $t$ while $\langle, \rangle_2$ is Euclidean. Finally, the spaces are oriented such that $\vol_V = \vol_1 \vol_2$. Let $\star_V$ be the Hodge star on $V$. The $\langle, \rangle_i$ induce Hodge stars $\star_i$ on $W_i$. For $\alpha_i, \beta_i \in W_i$, one finds
\begin{IEEEeqnarray}{rCl}
  \star_V (\alpha_1 \otimes \alpha_2) \wedge (\beta_1 \otimes \beta_2) &=& \langle \alpha_1, \beta_1 \rangle \langle \alpha_2, \beta_2 \rangle \vol_1 \vol_2 = (\star_1 \alpha_1 \wedge \beta_1) \wedge (\star_2 \alpha_2 \wedge \beta_2) \nonumber\\
  &=& (-1)^{k_1 (d_2 - k_2)} \star_1 \alpha_1 \wedge \star_2 \alpha_2 \wedge \beta_1 \wedge \beta_2.
\end{IEEEeqnarray}
Thus
\begin{equation}\label{eq:hodge_star_threading}
  \star_V (\alpha_1 \wedge \alpha_2) = (-1)^{k_1 (d_2 - k_2)} \star_1 \alpha_1 \wedge \star_2 \alpha_2.
\end{equation}

Simpler considerations lead to
\begin{IEEEeqnarray}{rCl}
  \star_{10A} \alpha &=& e^{(10-2k)A} \star_{10} \alpha, \qquad
  \star_{6A} \beta = e^{(4k-12)A} \star_6 \beta, \qquad
  \alpha \in \Omega^k(M_{10}), \beta \in \Omega^k(M_6).
\end{IEEEeqnarray}
With this in mind we turn to $\star_7$ where for $\alpha \in \Omega^k(M_6)$
\begin{IEEEeqnarray}{rCl}\label{eq:Hodge_Stars_7_and_6_related}
  \star_7 \alpha &=& \star_7 (\alpha \otimes \bI) = \eta \wedge \star_{6A} \alpha = e^{(4k-12)A} \eta \wedge \star_6 \alpha, \nonumber\\
  \star_7 (\alpha \otimes \eta) &=& e^{(4k-12)A} \star_{6} \alpha \otimes \star_1 \eta = (-1)^k e^{(4k-12)A} \star_6 \left[ \eta \lrcorner (\alpha \wedge \eta) \right].
\end{IEEEeqnarray}

\subsection{Dirac Algebra}
\label{sec:Fermionic_algebra_conventions}

We decompose the ten-dimensional Dirac matrices as
\begin{equation}
  \Gamma_\alpha = \sigma_1 \otimes \gamma_\alpha \otimes \bI, \qquad
  \Gamma_a = \sigma_2 \otimes \bI \otimes \gamma_a,
\end{equation}
where $\gamma_\alpha$ and $\gamma_a$ are Dirac matrices on $\AdS_3$ and $M_7$ respectively. For specific calculations, we use
\begin{IEEEeqnarray}{rCl}
  \gamma_0^{\AdS} &=& \imath \sigma_1, \qquad
  \gamma_1^{\AdS} = \sigma_2, \qquad
  \gamma_2^{\AdS} = \sigma_3,
\end{IEEEeqnarray}
as well as
\begin{IEEEeqnarray}{rClCrClCrCl}
  \gamma^{M_7}_1 &=& \sigma_1 \otimes \bI \otimes \bI, &\qquad&
  \gamma^{M_7}_2 &=& \sigma_2 \otimes \bI \otimes \bI, &\qquad&
  \gamma^{M_7}_3 &=& \sigma_3 \otimes \sigma_1 \otimes \bI, \nonumber\\
  \gamma^{M_7}_4 &=& \sigma_3 \otimes \sigma_2 \otimes \bI, &\qquad&
  \gamma^{M_7}_5 &=& \sigma_3 \otimes \sigma_3 \otimes \sigma_1, &\qquad&
  \gamma^{M_7}_6 &=& \sigma_3 \otimes \sigma_3 \otimes \sigma_2, \nonumber\\
  \gamma^{M_7}_7 &=& \sigma_3 \otimes \sigma_3 \otimes \sigma_3.
\end{IEEEeqnarray}
We have chosen the signs in $\gamma_0^{\AdS}$ and $\gamma_7^{M_7}$ such that
\begin{equation}
  \gamma^{\AdS}_{012} = -\bI, \qquad
  \gamma^{M_7}_{1234567} = -\imath.
\end{equation}
The ``intertwiner'' matrices are
\begin{IEEEeqnarray}{rCCCCCCl}
  C_{10} &=& \Gamma_{02468} &=& \sigma_2 \otimes \gamma_{02} \otimes \gamma_{246} &=& \sigma_2 \otimes C_3 \otimes M_7,
 \nonumber\\
  B_{10} &=& - \Gamma_{2468} &=& - \imath \sigma_3 \otimes \gamma^{\AdS}_2 \otimes \gamma^{M_7}_{246} &=& \imath \sigma_3 \otimes B_3 \otimes B_7
\end{IEEEeqnarray}
on $M_{10}$ and
\begin{equation}
  C_3 = \gamma^{\AdS}_{02}, \qquad
  C_7 = \gamma^{M_7}_{246}, \qquad
  B_3 = - \gamma^{\AdS}_2, \qquad
  B_7 = C_7
\end{equation}
for the internal manifolds. They satisfy
\begin{IEEEeqnarray}{rClCrCl}
  C_{10} \Gamma_A C_{10}^{-1} &=& (\Gamma_A)^T, &\qquad&
  B_{10} \Gamma_A B_{10}^{-1} &=& - (\Gamma_A)^*, \nonumber\\
  C_3 \gamma^{\AdS}_\alpha C_3^{-1} &=& - (\gamma^{\AdS}_\alpha)^T, &\qquad&
  B_3 \gamma^{\AdS}_\alpha B_3^{-1} &=& (\gamma^{\AdS}_\alpha)^*, \nonumber\\
  C_7 \gamma^{M_7}_a C_7^{-1} &=& - (\gamma^{M_7}_a)^T.
\end{IEEEeqnarray}

The chirality matrix is $\Gamma^{(10)} = \Gamma_{012\dots 9} = - \sigma_3 \otimes \bI \otimes \bI$ and thus that the chirality condition for IIB, $\Gamma^{(10)} \epsilon = -\epsilon$, reduces to $\sigma_3 \epsilon = \epsilon$. This also implies that
\begin{equation}
  \Gamma^{A_1 \dots A_5} \epsilon = \frac{1}{5!} \epsilon^{A_1 \dots A_5}_{\phantom{A_1 \dots A_5}B_1 \dots B_5} \Gamma^{B_1 \dots B_5} \epsilon.
\end{equation}
For a spinor of opposite chirality, a minus sign appears on the right hand side of the above equation.

\section{The Spin 2 Mass Operator}
\label{sec:appendix_spin_2}

We summarize the derivation of the mass operator of spin 2 fluctuations \cite{Gubser:1997yh,Constable:1999gb,Klebanov:2009kp,Bachas:2011xa}, following mainly \cite{Bachas:2011xa}. To begin we introduce the Lichnerowicz operator on a symmetric 2-tensor:
\begin{equation}
  \Delta_L h_{MN} \equiv - \nabla^P \nabla_P h_{MN} - [\nabla_M, \nabla^P] h_{NP} - [\nabla_N, \nabla^P] h_{MP}.
\end{equation}
Next we cite some standard results regarding perturbations of the metric:
\begin{IEEEeqnarray}{rCl}
  \delta R^K_{\phantom{K}LMN} &=& \nabla_M \delta \Gamma^K_{NL} - \nabla_N \delta \Gamma^K_{ML}, \nonumber\\
  \delta R_{MN} &=& \nabla_K \delta \Gamma^K_{NM} - \nabla_N \delta \Gamma^K_{KM}, \nonumber\\
  \delta \Gamma^K_{MN} &=& \frac{1}{2} g^{KL} \left( \nabla_M \delta g_{NL} + \nabla_N \delta g_{ML} - \nabla_L \delta g_{MN} \right).
\end{IEEEeqnarray}
These allow us to express the variation of the Ricci tensor in terms of the Lichnerowicz operator:
\begin{IEEEeqnarray}{rCl}
  \delta R_{MN} &=& \frac{1}{2} \left( - \nabla^P \nabla_P \delta g_{MN} + \nabla^P \nabla_M \delta g_{NP} + \nabla^P \nabla_N \delta g_{MP} - \nabla_M \nabla_N \delta g_P^{\phantom{P}P} \right) \nonumber\\
  &=& \frac{1}{2} \left( \Delta_L \delta g_{MN} + \nabla_M \nabla^P \delta g_{NP} + \nabla_N \nabla^P \delta g_{MP} \right).
\end{IEEEeqnarray}

As in the main text we denote the difference between the warped and unwarped ten-dimensional metric with a hat. That is, $d\hat{s}^2 = e^{2A} ds^2$. The Ricci tensors and scalars then satisfy
\begin{IEEEeqnarray}{rCl}
  \hat{R}_{MN} &=& R_{MN} - (\dim - 2) (\nabla_M \partial_N A - \partial_M A \partial_N A) + g_{MN} [\Delta A - (\dim - 2) dA^2], \nonumber\\
  \hat{R} &=& e^{-2A} [ R + 2 (\dim - 1) \Delta A - (\dim - 2)(\dim-1) dA^2],
\end{IEEEeqnarray}
where for our purposes $\dim = 10$.

At this stage we introduce a perturbation of the metric along the $\AdS_3$ factor:
\begin{equation}
  \delta \hat{g}_{\mu\nu} = e^{2A} \delta g_{\mu\nu} = e^{2A} h_{\mu\nu} (\AdS_3) Y_0 (M_7),
\end{equation}
where $h_{\mu\nu}$ satisfies transverse-traceless gauge conditions; $\nabla^\mu h_{\mu\nu} = h_\mu^{\phantom{\mu}\mu} = 0$. Note that this appendix uses greek indices to indicate $\AdS_3$ directions. Since $A$ depends only on the internal manifold, one finds that $\delta (dA)^2 = 0$, $\delta (\Delta_7 A) = g^{MN} \delta \Gamma_{MN}^k \partial_k A = - \frac{1}{2} h_\mu^{\phantom{\mu}\mu} g^{kl} \partial_k A \partial_l Y_0 = 0$ and $\delta (\nabla_\mu \partial_\nu A) = \frac{1}{2} h_{\mu\nu} g^{kl} \partial_k Y_0 \partial_l A$. It follows that $\delta \hat{R}$ vanishes, while
\begin{IEEEeqnarray}{rCl}
  \delta \hat{R}_{\mu\nu} &=& \frac{1}{2} \left[ Y_0 \Delta_L h_{\mu\nu} + h_{\mu\nu} \mathcal{L}_0 Y_0 \right] + h_{\mu\nu} Y_0 ( \Delta A - 8 dA^2 ).
\end{IEEEeqnarray}

Turning to the equation of motion,
\begin{equation}
  \hat{R}_{MN} - \frac{1}{2} \hat{g}_{MN} \hat{R} = \hat{T}_{MN},
\end{equation}
we need to consider the variation of the energy-momentum tensor. One could evaluate $\delta \hat{T}_{\mu\nu}$ explicitly for the background at hand, yet it has been argued in \cite{Bachas:2011xa} that this is not necessary. At the linearized level the spin 2 modes are decoupled from the rest of the spectrum. Due to the symmetries of the background it follows then that
\begin{equation}
  \delta \hat{T}_{\mu\nu} = \frac{1}{3} \delta \hat{g}_{\mu\nu} \hat{T}_\rho^{\phantom{\rho}\rho}.
\end{equation}
Using $R_{\kappa\lambda\mu\nu} = - (g_{\kappa\mu} g_{\lambda\nu} - g_{\kappa\nu} g_{\lambda\mu})$ one calculates $\hat{T}_\rho^{\phantom{\rho}\rho} = - 6 e^{-2A} + 3 e^{-2A} (\Delta A - 8 dA^2) - \frac{3}{2} \hat{R}$ and thus
\begin{equation}
  \delta \hat{T}_{\mu\nu} = \delta g_{\mu\nu} \left(- 2 + \Delta A - 8 dA^2 - \frac{1}{2} e^{2A} \hat{R} \right).
\end{equation}
Combining our previous results,
\begin{IEEEeqnarray}{rCl}
  \delta \left( \hat{R}_{\mu\nu} - \frac{1}{2} \hat{g}_{\mu\nu} \hat{R} \right) = \frac{1}{2} \left[ Y_0 \Delta_L h_{\mu\nu} + h_{\mu\nu} \mathcal{L}_0 Y_0 - e^{2A} \hat{R} \right] + h_{\mu\nu} Y_0 (\Delta_7 A - 8dA^2),
\end{IEEEeqnarray}
and we can conclude that
\begin{equation}
  0 = Y_0 (\Delta_L + 2) h_{\mu\nu} + h_{\mu\nu} \mathcal{L}_0 Y_0,
\end{equation}
the equation of motion for a spin 2 fluctuation of mass $\mathcal{L}_0$.

\section{Example: Products of K\"ahler\--Ein\-stein Spaces}
\label{sec:product_of_kahler_einstein}

A simple yet interesting class of solutions that has been discussed in section 6 of \cite{Gauntlett:2006ns} arises if $M_6$ is the product of K\"ahler-Einstein spaces. The D3-D3 intersection that we discussed briefly in section \ref{sec:Example_D3D3} falls into this class. Following \cite{Gauntlett:2006ns}, one uses the ansatz $M_6 = KE_2^{(1)} \times KE_2^{(2)} \times KE_2^{(3)}$. The metric is simply
\begin{equation}
  ds_6^2 = \sum_i ds^2(KE_2^{(i)})
\end{equation}
and since each factor is Einstein the Ricci form decomposes as $\rho = \sum_i l_i J_i$ for some constants $l_i$. It follows that $R = 2 \sum_i l_i$. In general the $K_2^{(i)}$ are of dimension two, yet by considering the special case for which two $l_i$ are equal the analysis of \cite{Gauntlett:2006ns} includes the case of $M_6 = KE_4 \times KE_2$. The curvature constraint \eqref{eq:M6_curvature_constraint} is solved if $l_1 l_2 + l_2 l_3 + l_3 l_1 = 0$. The flux \eqref{eq:F5} is
\begin{equation}
  F = \frac{1}{2}
  \begin{pmatrix}
    \left( 1 - \frac{1}{4} e^{4A} l_1 \right) J_1 & 0 & 0 \\
    0 & \left( 1 - \frac{1}{4} e^{4A} l_2 \right) J_2 & 0 \\
    0 & 0 & \left( 1 - \frac{1}{4} e^{4A} l_3 \right) J_3
  \end{pmatrix}.
\end{equation}

Let us revisit the $AdS_3 \times S^3 \times T^4$ solution of section \ref{sec:Example_D3D3}. Here, $M_6 = \mathds{CP}^1 \times T^4$ and thus $l_2 = l_3 = 0$, $l_1 = 1$. Therefore $R = 2$ and $e^{4A} = 4$. Studying the constraint equation for one-form wave functions in section \ref{sec:betti_multiplets}, we find that for one-forms $Y_1$ along $\mathds{CP}^1$ $y_1 = 0$. Along $T^4$ on the other hand $y_1 = \pm\frac{1}{8}$. Of course one should keep in mind that $H^1(\mathds{CP}^1) = 0$.

A similar example is $M_6 = H^2 \times KE_4^+$. Here $KE_4^+$ is a K\"ahler-Einstein manifold of positive curvature such as $\mathds{CP}^1 \times \mathds{CP}^1$, $\mathds{CP}^2$ or a del Pezzo surface $dP_k$, $k = 3, \dots, 8$. Here $l_1 = -1$ and $l_2 = l_3 = 2$. Then $R = 6$ and $e^{4A} = \frac{4}{3}$. $Y_1 \in \Omega^1 (H^2)$ results in $y_1 = \pm \frac{1}{2}$, $Y_1 \in \Omega^1(KE_4^+)$ on the other hand in $y_1 = \pm \frac{1}{8}$.

\bibliographystyle{ytphys}
\small\baselineskip=.97\baselineskip
\bibliography{ref}

\end{document}